\documentclass[mathleft
]{an}
\usepackage{graphicx}
\usepackage{times}
\newcommand{\clfort}{{RXJ1415.1+3612}}
\newcommand{\oii}{\mbox{[O\,{\scriptsize II}]}}

\begin{document}

\Pagespan{1}{}
\Yearpublication{2009}%
\Yearsubmission{2009}%
\Month{11}%
\Volume{999}%
\Issue{99}%

\title{The Evolution of Cluster Early-Type Galaxies over the Past 8 Gyr}

\author{Alexander Fritz\thanks{Corresponding author:
  afritz@gemini.edu \newline} \and Inger J{\o}rgensen
  \and Ricardo P. Schiavon \and Kristin Chiboucas}
  
\titlerunning{The Evolution of Cluster Early-Type Galaxies over the Past 8 Gyr}
\authorrunning{A. Fritz et al.}
\institute{Gemini Observatory, 670 N.\ A'ohoku Place,
Hilo, HI 96720, USA}

\received{\today}


\abstract{%
We present the Fundamental Plane (FP) of early-type galaxies in the clusters of 
galaxies \clfort\ at $z=1.013$. This is the first detailed FP investigation of
cluster early-type galaxies at redshift $z=1$. The distant cluster galaxies follow
a steeper FP relation compared to the local FP. The change in the slope of the FP 
can be interpreted as a mass-dependent evolution. To analyse in more detail 
the galaxy population in high redshift galaxy clusters at $0.8<z<1$, we
combine our sample with a previous detailed spectroscopic study of 38
early-type galaxies in two distant galaxy clusters, RXJ0152.7-1357 at $z=0.83$
and RXJ1226.9+3332 at $z=0.89$. For all clusters Gemini/GMOS spectroscopy
with high signal-to-noise and intermediate-resolution has been acquired to
measure the internal kinematics and stellar populations of the galaxies.
From HST/ACS imaging, surface brightness profiles, morphologies and structural
parameters were derived for the galaxy sample.
The least massive galaxies ($M=2\times10^{10}M_{\sun}$) in our sample have
experienced their most recent major star formation burst at
$z_{\rm form}\sim 1.1$. For massive galaxies ($M>2\times10^{11}M_{\sun}$) the
bulk of their stellar populations have been formed earlier $z_{\rm form}\ga 1.6$.
Our results confirm previous findings by J\o rgensen et al.
This suggests that the less massive galaxies in the distant clusters have
much younger stellar populations than their more massive counterparts.
One explanation is that low-mass cluster galaxies have experienced more
extended star formation histories with more frequent bursts of star formation
with shorter duration compared to the formation history of high-mass cluster
galaxies.}

\maketitle

\section{Introduction}

Clusters of galaxies at intermediate-redshift ($1<z<2$) provide extremely useful
laboratories to map the large-scale structure of the universe and study the
formation and evolution of cluster galaxies.

There are only a small number of clusters of galaxies known at
redshifts $z>1$ and our understanding of the galaxy population in these high
redshift clusters is limited (e.g., Rosati et al. 2004; Demarco et al. 2007;
Lidman et al. 2008; Mei et al. 2009). $X$-ray surveys (e.g., Stanford et al. 2001) have
detected about a dozen clusters with $z\ge 1$. Current surveys carried out in
the $X$-rays (Romer et al.\ 2001; Stanford et al. 2006; Finoguenov et al. 2007),
optical (e.g., Gladders \& Yee 2005), infrared (Stanford et al.\ 2005;
Lawrence et al.\ 2007), and radio (Bran\-chesi et al.\ 2006)
provide the basis for new but small number of discoveries. Upcoming surveys
using the Sunyaev-Zel'dovich effect are expected to discover between 100 to
1000 new massive galaxy clusters at high-redshifts of $z>1$
(Carlstrom et al.\ 2002). 

In this context, new spectroscopic observations of clusters of galaxies play a
key role for understanding galaxy evolution in dense environments. 
Previous studies at $z\sim1$ concentrated on a few of the brightest (hence more
massive $\ga 2\times10^{11}\,M_{\sun}$) early-type cluster members
(van Dokkum \&  Stanford 2003; Holden et al.\ 2005), because high
signal-to-noise ($S/N$) spectroscopy of distant galaxies is very expensive in
telescope time ($\ga$20 hours at 8-10m facilities). In particular, these two
Fundamental Plane analyses were restricted to three and four brightest early-type
cluster members, respectively.

At redshifts of $z\sim1.5$, both cluster (e.g., Cucciati et al. 2006;
Cooper et al. 2007) and field (e.g., Bell et al. 2004; Faber et al. 2007;
Ferreras et al. 2009; Pozzetti et al. 2009) surveys show a significant
increase (by about a factor of two) in the number density of 
early-type galaxies from high\-er redshift to the present day. In other
words, only about 50\% of the total baryonic (stellar) mass in early-type
galaxies has been generated before $z_{\rm form}\sim1.5$, whereas the
remainder of the bulk of the stars has been assembled at much later epochs
of the universe $z_{\rm form}\la 2$. 

Through an investigation of the Fundamental Plane constraints on the
evolution and formation history of early-type galaxies as well as their dark
matter content are possible. The Fundamental Plane (FP) is a tight linear relation in
three-dimensional log-space defined by the effective (half-light) radius
$r_{{\rm e}}$, the average surface brightness within $r_{{\rm e}}$
($\langle I_{{\rm e}}\rangle$) and the central velocity dispersion ($\sigma$)
of early-type galaxies (Djorgovski \& Davis 1987; Dressler et al.\ 1987;  
J\o rgensen et al.\ 1996, hereafter JFK96). This scaling relation has proven to
be a powerful tool in measuring the luminosity-weighted average ages of
early-type galaxies, \newline both at local (e.g., Bender et al.\ 1992;
JFK96; Bernardi et al.\ 2003) and at intermediate up to redshifts of $z\sim1.2$
(e.g., J\o rgensen et al.\ 1999; Wuyts et al.\ 2004; Fritz et al.\ 2005;
di Serego Alighieri et al. 2005; Treu et al.\ 2005; J\o rgensen et al.\ 2006,
2007; Fritz et al.\ 2009). The mean age of the stellar populations in these
galaxies can be directly probed, because an evolution in the zero-point offset
of the FP with increasing redshift can be directly related to a change in the
average mass-to-light ($M/L$) ratio of the galaxies under consideration. 

At $z\sim1$ our understanding of the FP is very limited. Previous works based on
small number statistics at $z\sim1.3$ found a FP relation with a large intrinsic
scatter, twice as high as in the Coma cluster (van Dokkum \&  Stanford 2003;
Holden et al.\ 2005), with the properties of these early-type cluster galaxies
poorly understood. This larger intrinsic scatter is in disagreement with the
FP for two galaxy clusters at $z=0.8-0.9$ (J\o rgensen et al.\ 2006, 2007).
In this high redshift domain selection effects play an important role. Therefore,
both large, well defined sample sizes as well as high quality data are needed
to separate observational limitations (like sample selection or cosmic variance)
from the underlying physical processes in these distant galaxies (see also
discussion in Fritz et al. 2009).

We present the FP for 12 early-type galaxies at $z=1.013$ that are associated 
with the rich, massive and $X$-ray luminous cluster of galaxies \clfort. 
The galaxy cluster has been observed using a combination of
high $S/N$ GMOS spectroscopy at the 8m Gemini North telescope and deep imaging
of the Advanced Camera for Surveys (ACS) onboard Hubble Space Telescope (HST).
Our most distant sample reaches an absolute magnitude limit of $M_B\sim -21$ mag
(rest-frame), which makes this work less susceptible to selection effects.
Our study of \clfort\  is part of the Gemini/HST Galaxy Cluster Project
(J\o rgensen et al.\ 2005), an extensive observational program to investigate
the evolution, star formation and chemical enrichment history of galaxies in
rich galaxy clusters from $z=1$ to the present-day universe. Previous results
on the Gemini/HST Galaxy Cluster Pro\-ject have been published on
the stellar populations of the cluster RXJ0152.7-1357 at $z=0.83$
(J\o rgensen et al.\ 2005), the stellar content and FP of RXJ0142.0+2131
at $z$$=$$0.28$ (Barr et al. 2005, 2006) and the FP of RXJ0152.7-1357 and 
RXJ1226.9+3332 at $z=0.89$ (J\o rgensen et al.\ 2006, 2007). Throughout this
article we adopt a $\Lambda$CDM cosmology for a flat low-density Universe with
$H_0$$=$$70\,{\rm km\,s^{-1}}$ ${\rm Mpc^{-1}}$, $\Omega_M=0.3$, and
$\Omega_{\rm \Lambda}=0.7$. Unless otherwise noted, all magnitudes are given
in the Vega system.

\section{The Gemini/HST Galaxy Cluster Project}

The Gemini/HST Galaxy Cluster Project (PI: I. J\o rgensen) is an extensive
observational campaign to investigate the formation and evolution of distant
galaxies in rich galaxy clusters from $z=1$ to the present-day
(J\o rgensen et al.\ 2005).
For 15 massive, $X$-ray luminous selected galaxy clu\-sters from $z=0.1$ to 1, optical
intermediate-resolu\-tion Gemini/GMOS spectroscopy and HST/ACS$+$WFPC2
ima\-ging have been acquired. These rich galaxy clusters have $X$-ray
luminosities in the order of
$L_X\,{\rm (0.1-2.4\,keV)}=2\times10^{44}$ ergs s$^{-1}$, which in particular
range from about $0.5-1$ times the $X$-ray luminosity of the Coma
galaxy cluster. The high $S/N$ and moderately high resolution 
($\langle S/N\rangle \sim 25\,\AA^{-1}$ in the rest-frame) in the GMOS galaxy
spectra  allows us to study in detail the internal kinematics and stellar
pop\-ulations of the galaxies.
The nearby Coma, Perseus and \newline Abell~4038 galaxy clusters at $z=0.02$ are used as
local comparison samples (J\o rgensen et al.\ 1995; 2005; JFK96). The properties and
selection effects of this sample have \newline been extensively tes\-ted and therefore
are well understood (JFK96; J{\o}rgensen 1999).

The main science drivers and objectives of the galaxy cluster project are
(see also partly J\o rgensen et al.\ 2005):
\begin{itemize}
\item Investigate the star formation history (SFH) of cluster galaxies as a
function of redshift;
\item  Study the early-type galaxy population over a wide mass range at
$z=1$;
\item Construct scaling relations and study the evolution of their
zero-point, slope and scatter;
\item Test the evolution for different galaxy morphologies;
\item Measure absorption line strengths to constrain the evolution
of the stellar populations and the overall chemistry in the galaxies;
\item Constrain the chemical enrichment history of the galaxies by computing
ages, metallicities and abundance ratios; 
\item Test the model predictions for possible variations in the initial mass
function (IMF);
\item Search for possible dynamical and morphological
substructure of the clusters under investigation.
\end{itemize}
The combined results for all of these approaches will enable a detailed picture
of the involved formation processes and subsequent evolution of galaxies over
the past 8 Gyr.

\section{\clfort: Observations and Data}

\subsection{GMOS Spectroscopy}

Target galaxies were selected using colour-colour and \newline colour-magnitude diagrams based on
deep ground-based \newline $r'i'z'$  GMOS-N imaging of the central $5\farcm5\times5\farcm5$ 
field of the rich clusters of galaxies \clfort\ at $z=1.013$ 
down to $z_{850}<23.87$ mag (J\o rgensen et al.\ 2009, in prep). 
Follow-up MOS spectroscopy of 37 galaxies in the field of \clfort\ was acquired with
the Gemini Multi-Object Spectrograph (GMOS-N, Hook et al.\ 2004) at Gemini North
in queue mode in dark time in 2005 as part of Gemini program GN-2005A-Q-33.
The imaging data for \clfort\ were observed as part of Gemini progra\-ms
GN-2003A-DD-4 and GN-2003A-SV-80.
Spe\-ctroscopic observations made use of the unique nod-and-shuffle (N\&S)
GMOS mode at Gemini (Glazebrook \& Bland-Hawthorn 2001) to reach both
the magnitude limit and to avoid limitations of sky-subtraction systematics.
Briefly, the N\&S observing technique allows an improved subtraction of strong
telluric night sky emission lines for faint targets by using telescope (nod) and
detector charge shifts (shuffle), while reading out the detector only after 30
minutes of integration. A series of 48 individual exposures yielded a total
integration time of 24 hours. Adopting an instrument setup with slit widths
of $1''$ 
and the R400 grating results in a resolving power of $R\sim2000$ and an
instrument resolution in the final spectra of
$\sigma \sim 105\ {\rm km\,s^{-1}}$ at 4300\AA, which is sufficient to measure
both internal kinematics (velocity dispersions) and
absorption line-strengths of the galaxies. The longpass filter
{\tt OG515\_G0306} was used to avoid second order contamination of the spectra.

Radial velocities and velocity dispersions ($\sigma$) of the galaxies were
measured using a penalized maximum-likeliho\-od fitting algorithm in pixel space
(Gebhardt et al. 2003). To ensure a reliable error treatment and test the impact of
systematic effects, J\o rgensen et al.\ (2005) performed Monte Carlo simulations
for artificially generated galaxy spectra and different stellar templates.
Results showed that the $S/N$ of the galaxy spectra does not significantly
affect the systematic error. Galaxies with low velocity dispersions\newline
(${\rm log}\ \sigma\le1.74$), which is 1/2 times the instrumental resolution
might be subjected to systematic effects and therefore have been excluded from
the analysis of the scaling relations. The spectroscopy of  \clfort\ will be
published in J\o rgensen et al.\ (2009).

The GMOS N\&S observations for \clfort\ \newline yielded in a total
spectroscopic sample of $N_{\rm spec}=37$ galaxies. Out of this sample, 18
galaxies are cluster members, of which 14 galaxies are within the ACS
field-of-view of the galaxy cluster. For two galaxies out of these 14 inside the
ACS field, no velocity dispersion could be derived due to low $S/N$ of those
spectra.

\begin{figure}
\includegraphics[width=0.46\textwidth]{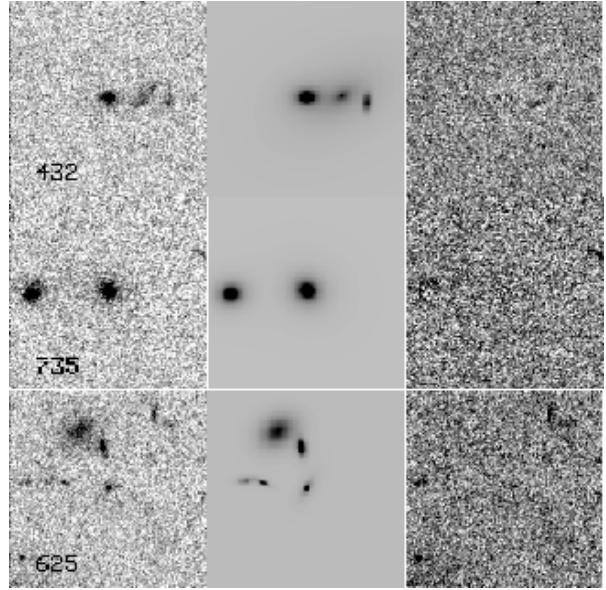}
\caption{\label{Fig1} HST/ACS surface brightness modelling for three cluster galaxies
in \clfort\ at $z=1.02$. The examples show the results of $r^{1/4}$-law models for
three different cluster members over a wide range in magnitudes and
sub-structure. Close neighbouring galaxies were independently modelled and
then subtracted. Left panel: Distortion corrected ACS original image,
middle panel: Adopted model with neighbouring objects that are subtracted out,
Right panel: Residual image.}
\end{figure}

\begin{figure*}
\includegraphics[width=0.36\textwidth,angle=-90]{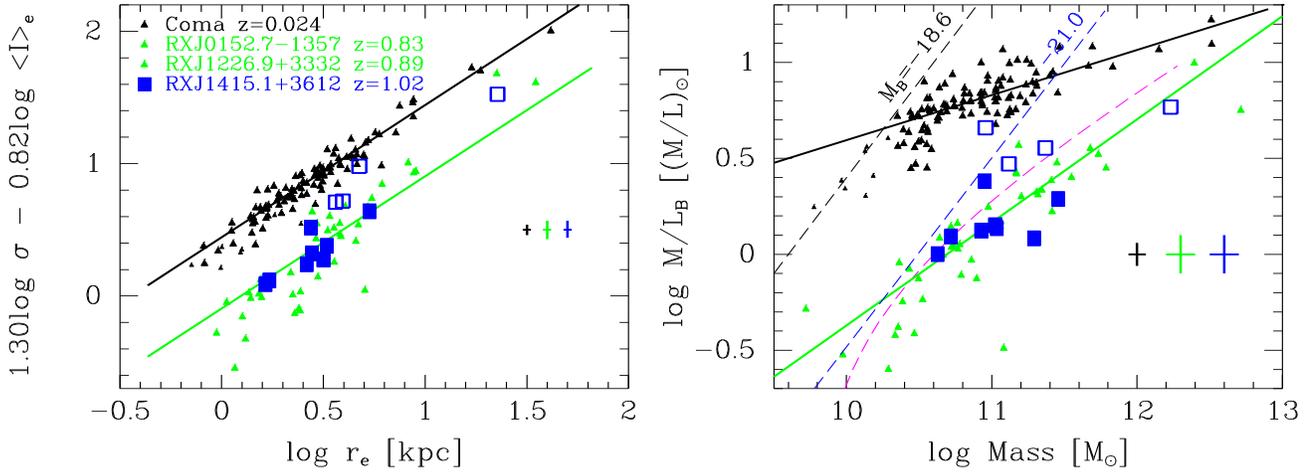}
\caption{\label{Fig2} 
Fundamental Plane (FP) of early-type galaxies in RXJ0152.7-1357 ($z=0.83$, green
triangles), RXJ1226.9+3332 ($z=0.89$, green triangles), both from J\o rgensen et al.\ (2006, 2007), and
RXJ1415.1+3612 ($z=1.02$, blue squares). Left: Edge-on projection
of FP. The Coma cluster is used as a local reference (black triangles, J{\o}rgensen 1999).
The solid black line represents the fit to the local Coma Cluster sample. The
solid green line represents the fit to the Coma Cluster sample offset to the
median zero-point of the two distant clusters at $z=0.8-0.9$.
Open symbols are galaxies with emission lines. Right: $M/L-M$ plane. Dashed
lines indicate the magnitude limits of Coma (black) and the distant cluster at
$z=1$ (blue). Solid green line is the fit to the two clusters at $z=0.8-0.9$.
Internal uncertainties for the local reference, the two clusters at $z=0.8$ and
the highest redshift cluster at $z=1$ are given as representative error bars.
We find a mass-dependent evolution, where less massive galaxies show signatures
of recent star formation episodes since $z\sim1.1-1.3$. The dashed, magenta
line shows the predictions of the single-burst stellar population models by
Thomas et. al. (2005).}
\end{figure*}

\subsection{HST/ACS Photometry}

We use deep HST/ACS imaging for \clfort\ from the HST archive which was acquired
during Cycle 14 (GO 10496, PI: S. Perlmutter). The primary aim of this program
is to search for "dust free" supernovae of Type Ia (SNe Ia) in high-redshift
clusters to constrain dark energy time variations. 
Each of the eight individual pointings in the
F850LP filter was split into four exposure times of 340~s to 375~s, yielding a
total integration time of 9920~s. The observations in the F775W filter will not
be used in the following.
The HST/ACS surface brightness distribution of 30 galaxies within the ACS
pointing of the galaxy cluster was modelled 
using the publicly available 2D fitting algorithm GALFIT (Peng et al.\ 2002).
Each galaxy surface brightness profile was separately analysed with
seeing-convolved, sky-corrected model profiles. ACS structural parameters
(total magnitude, effective (half-light) radius $r_{{\rm e}}$, and average
surface brightness within $r_{{\rm e}}$, $\langle I_{{\rm e}}\rangle$)
were derived by fitting a pure de Vaucouleurs ($r^{1/4}$) profile to the
luminosity distribution of the galaxy of interest. Afterwards the derived
photometry were calibrated to rest-frame Johnson $B$-band using our $r'i'z'$
GMOS photometry and also with evolutionary synthesis models
(see J\o rgensen et al.\ 2005 for details). We also tested the
effects of other surface brightness models (e.g. single Sersic fits with varying
$n$-parameter) on the structural parameters. The results remained very stable for
our galaxies. In particular, the combination of the FP parameter that
enters the FP, (${\rm log}\,r_e+\beta\,{\rm log}\,\langle I_{{\rm e}}\rangle$ with
$\beta$ in the range $0.7<\beta<0.8$), is highly stable and differs only
little for the choice of the profile (Fritz et al. 2005; Chiboucas et al. 2009).
Elliptical (E) and S0 galaxies were morphologically classified using a visual
inspection, the profile fitting results, and their bulge-to-total fractions. For
the following analysis, however, the early-type galaxies are treated as one
homogenous group and no separation into the different morphological sub-classes
has been made. To account for point-spread function (PSF) variation and
distortion effects, a unique PSF was generated for each galaxy according to its
position on the ACS chip using the TinyTim version 6.2 package (Krist 1995).
Similar results are found when, instead of artifically created PSF stars, 
real star measurements are used (Chiboucas et al. 2009).

In Fig.~\ref{Fig1}  we show the distortion corrected ACS image of three
different cluster members over a wide range in magnitudes and sub-structure 
along with the adopted model and residual images. Close neighbouring galaxies
were independently modelled and then subtracted. Fig.~\ref{Fig1} gives the
results of our surface brightness modelling using $r^{1/4}$-law profiles to
model the luminosity distribution of our sample galaxies.

\section{Results}

\subsection{A detailed FP of Cluster E+S0 galaxies at $z=1$}

As a first step we construct the FP for cluster E+S0 galaxies at $z=1$. 
Fig.~\ref{Fig2} shows the FP for the cluster E+S0 galaxies in the
rest-frame Johnson $B$-band, compared to 116 early-type
galaxies in the Coma cluster (J{\o}rgensen 1999). The local reference is
indicated with small triangles, whereas the distant E+S0 galaxies in 
\clfort\ at $z=1$ are displayed as the large squares. The two galaxy clusters 
RXJ0152.7-1357 and RXJ1226.9+3332 at $z=0.8-0.9$, both from
J\o rgensen et al.\ (2006, 2007), are denoted as green triangles. Filled
symbols denote E+S0 cluster members without emission lines, open symbols
cluster galaxies with emission lines.

To limit differences in the sample selection of the Coma cluster and the
high-redshift sample, low-mass galaxies \newline with $M<10^{10.3}M_{\sun}$ as well as
emission line galaxies were excluded. The FP of the Coma cluster
(J{\o}rgensen et al. 2006) is given by:
\begin{equation}
{\rm log}\,r_e=(1.30\pm0.08){\rm log}\,\sigma-(0.82\pm0.03){\rm log}\,
\langle I_{{\rm e}}\rangle+\gamma,
\end{equation}
with $\gamma=-0.443$, denoted in Fig.~\ref{Fig2} by the black solid line.

Fig.~\ref{Fig2} presents the most detailed ever FP of cluster E+S0 galaxies at
$z=1$. The FP for galaxies at $0.8<z<1.0$ has a different slope than the local Coma
FP. The distant FP for the high-redshift sample is not only offset from the local
reference, but appears to have a steeper slope. The slope difference between the
distant and the local relation is a strong indication of a 
mass-dependent evolution with a stronger evolution for less massive galaxies
(see below). The results for the highest redshift cluster confirm previous
findings for the two lower redshift clusters at $z=0.8-0.9$ by
J\o rgensen et al.\ (2006, 2007). Interestingly, \clfort\ cluster
galaxies at $z=1$ with significant \oii\,3727
emission lines (defined with equivalent widths ${\rm EW}\ge-5 \AA$; Balogh et
al. 1997) are offset from the galaxies without emission lines and display a
larger scatter. The galaxies with emission lines show higher $M/L$ ratios than
non-emission line galaxies, a trend which is opposite than expected. Partly
this can be explained that the star formation in these galaxies is weak and
that the net effect in the galaxy luminosity is small. Interestingly, also the
Brightest Cluster Galaxy (BCG) exhibits weak star formation but follows the same
FP relation as defined by the non-emission cluster members. In case of
the BCG, the measured recent star formation activity could be explained with a
`frosting' scenario that does not move the galaxy away from the FP of the
cluster galaxies at $z=1$. The location of the other galaxies with emission
lines is still an open question and will be investigated further in
J\o rgensen et al.\ (2009).

The observed evolution of the $M/L$ ratio as derived from the FP
depends on the age of the stellar population of the galaxies under consideration
(e.g., Fritz et al. 2009). As the FP relation can be translated into a
relationship between the $M/L$ ratio and the stellar mass of a galaxy, an
evolution in the FP zero-point offset from the local FP relation with
increasing redshift can be directly related to a change in the average $M/L$
ratio of the galaxies (Djorgovski \& Santiago 1993). 

The distant cluster galaxies follow a steeper $M/L$-mass relation than 
the local reference (see Fig.~\ref{Fig2}). This slope change can be interpreted
as a different dependence of the epoch of the most recent major star formation
episode with the stellar mass of a galaxy. 
Less massive galaxies ($M\sim2\times10^{10}M_{\sun}$) have experienced their
last major star formation burst at $z_{\rm form}\sim 1.1$, whereas for
massive galaxies ($M>2\times10^{11}M_{\sun}$) the majority of their stellar
populations have been formed earlier $z_{\rm form}\ga 1.6$. Our results
confirm previous findings by J\o rgensen et al.\ (2006, 2007) for  
RXJ0152.7-1357 and RXJ1226.9+3332 at $z=0.8-0.9$.

Using absorption-line strength data of nearby early-type galaxies in
high-density environments, Thomas et al. (2005) predicted the SFHs of these
galaxies for different galaxy masses. This single-burst stellar population (SSP) model
is indicated as the dashed magenta line in Fig.~\ref{Fig2}. Both the relative
timing of star formation ($t_{\rm form}< 11$ Gyr) and the slope change for
less massive galaxies are consistent with our findings for the high-redshift 
clusters.

This result can be reconciled with the \textit{down-sizing} formation scenario
(Cowie et al. 1996). Massive galaxies are dominated by red old passively evolving stellar
populations, whereas less massive systems have more extended SFH
and continue to form stars at much later epochs. Both the mass
assembly and SF are accelerated in massive systems in high density
environments, whereas processes work on longer timescales in less massive (smaller)
systems (Fritz 2007). Recent similar results derived for field E+S0 galaxies up
to $z\sim 1$ suggest that there is only a weak dependence of the overall galaxy
properties with their environment (Fritz 2007; Fritz et al. 2009).
The formation epoch of the field E+S0 galaxies depends mainly on their mass but
only wea\-kly on their luminosity, whereas the underlying environment regulates the
time scales of the most recent star formation episodes. This gets support by
similar results of a photometric study of cluster early-type galaxies at
$z=1.24$ (Rettura et al. 2008).

\section{Summary and Main Conclusions}

Galaxies in clusters at intermediate redshift are useful
pro\-bes to study the formation and the evolution of early-type galaxies.
By measuring the kinematics and structure parameters of these galaxies
their evolution in mass and mass-to-light ($M/L$) ratios can be derived.
This allows us to put constraints on the formation epoch and
subsequent evolution of spheroidal galaxies up to the present-day
as well as to critically test the models of galaxy formation and evolution.

In this paper we have presented the results of a detailed study of 
early-type galaxies in the distant clusters of galaxies \clfort\ at $z=1.013$.
Based on HST/ACS imaging, surface brightness profiles, morphologies and
stru\-ctural parameters were derived for 30 galaxies within the HST/ACS
field-of-view of the galaxy cluster. We have constructed the first detailed
Fundamental Plane (FP) of 12 clu\-ster early-type galaxies at redshift $z=1$. This
analysis is based on a sample size that represents a factor of 3 or more
improvement compared to previous studies. We have combined our distant cluster
sample with our previous detailed spectroscopic study of 38 early-type galaxies
in two distant galaxy clusters at $z=0.8-0.9$, RXJ0152.7-1357 and
RXJ1226.9+3332 (J\o rgensen et al.\ 2006, 2007). This allows us to
study in more detail the early-type galaxy population in high redshift galaxy
clusters up to redshift unity. Compared to the local reference, the distant
cluster galaxies follow a steeper FP relation and $M/L$-mass relation. This
slope change can be interpreted as a difference in the epoch of the
last major star formation episode for galaxies with different stellar masses.
The least massive galaxies ($M=2\times10^{10}M_{\sun}$) in our sample have
experienced their most recent major star formation burst at $z_{\rm form}\sim 1.1$,
whereas for massive galaxies ($M>2\times10^{11}M_{\sun}$) the majority of their
stellar populations have been formed earlier $z_{\rm form}\ga 1.6$.
The dependence of the formation epoch on the mass confirms previous results
by J\o rgensen et al.\ (2006, 2007).

A similar mass-dependent evolution with
more extended SFH has recently been found for field early-type galaxies
(Fritz et al. 2009). Our results can be understood if less massive cluster
galaxies have on average younger stellar populations. The distant galaxy
population in these low-mass galaxies could have been built up over longer time
scales and over more extended star formation histories of shorter duration but
more frequent star formation bursts, compared to the formation history of the
high-mass galaxy cluster counterparts.



Our results support a galaxy formation scenario according to 
the \textit{down-sizing} picture (Cowie et al. 1996), where the mass of galaxies
hosting star formation processes decreases with the age of the Universe. 
A combined analysis of the FP and the absorption line indices for
\clfort\ as well as RXJ0152.7-1357 and RXJ1226.9+3332 will be presented in
J\o rgensen et al.\ (2009).

\acknowledgements
AF and IJ acknowledge support from grant HST-GO-10826.01 from STScI. 
STScI is operated by AURA, Inc., under NASA contract NAS 5-26555.
Based on observations obtained at the Gemini Observatory
under Gemini programs GN-2002B-SV-90, GN-2002B-Q-29, GN-2003A-SV-80,
GN-2003A-DD-4, GN-2004A-Q-45, and GN-2005A-Q-33,
which is operated by AURA, Inc., under cooperative agreement with the NSF,
on behalf the Gemini Partnership: the NSF, STFC (UK), NRC (Canada),
CONICYT (Chile), ARC (Australia), CNPq (Brazil), and SECYT (Argentina).


\end{document}